\documentclass[12pt]{article}

\usepackage[hmargin=2.5cm,vmargin=3cm]{geometry}
\usepackage{graphicx}
\usepackage{amsmath}
\usepackage{wasysym}
\usepackage{enumerate}
\usepackage{graphicx}  
\usepackage{dcolumn}   
\usepackage{bm}        
\usepackage{amssymb}   
\usepackage{enumerate}
\usepackage{amsmath}
\usepackage{float}
\usepackage{multibib}
\usepackage{cite}
\newcommand{\bey}{\begin{eqnarray}}
\newcommand{\eey}{\end{eqnarray}}

\begin{document}
\title{ A post-selected quantum model of cosmic acceleration}
  \author {Dimitris Lionas\footnote{dlionas@ac.upatras.gr},   Charis Anastopoulos\footnote{anastop@upatras.gr}, and Konstantinos N. Gourgouliatos\footnote{kngourg@upatras.gr} \\
 {\small Laboratory of Universe Sciences, Department of Physics, University of Patras, 26500 Greece} }
\maketitle

\begin{abstract}
The origin of cosmic acceleration remains a central problem in cosmology, commonly attributed to a cosmological constant within the $\Lambda$CDM model or to dynamical dark energy. Here, we develop an alternative approach in which acceleration emerges from quantum post-selection, a standard feature of quantum theory that is not usually incorporated into cosmological modelling. While quantum theory admits both pre-selected and post-selected ensembles, quantum cosmological models are almost exclusively formulated in terms of initial conditions. Building on previous work on post-selected quasiclassical dynamics, we construct a minimal predictive cosmological model in which post-selection and coarse-graining generate effective late-time acceleration without introducing a cosmological constant, dark energy, or modifications of general relativity. The resulting expansion history is highly constrained theoretically and depends on at most two parameters beyond standard Friedmann evolution. Confrontation with type Ia supernova and cosmic chronometer data yields statistically competitive fits while naturally avoiding the coincidence problem. The model also reproduces the standard radiation- and matter-dominated behaviour at early times and predicts a present-day jerk parameter significantly different from the $\Lambda$CDM value. These results suggest that cosmic acceleration may arise as a macroscopic quantum cosmological effect rather than from additional cosmological fluids or modified gravitational dynamics.

\end{abstract}

\newpage
The origin of cosmic acceleration \cite{Riess98, Perl99} remains one of the central open problems in cosmology. Current observations are accurately described by the $\Lambda$CDM model, in which acceleration is attributed to a cosmological constant. Despite its empirical success, the physical origin of the accelerated expansion remains unclear, motivating the search for alternative mechanisms.

In quantum theory, probabilities may be conditioned on both initial and final states. While pre- and post-selected ensembles are treated symmetrically at the microscopic level \cite{ABL}, cosmological models are almost exclusively formulated in terms of initial conditions alone. However, recent work showed that quantum post-selection, combined with coarse-graining, generically modifies effective quasiclassical dynamics, leading to departures from standard Hamiltonian evolution \cite{Ana25}.  

Here, we develop the first observationally testable cosmological realization of this mechanism. The resulting model, which we term post-selected quantum cosmology (POQCO), predicts late-time accelerated expansion without introducing a cosmological constant, dark energy, or modifications of general relativity. Importantly, the effective dynamics is largely insensitive to the details of the underlying quantum theory, leading to a highly constrained expansion history with at most two additional phenomenological parameters beyond standard FRW cosmology.

We confront the model with type Ia supernova and cosmic chronometer observations and find statistically competitive fits despite its restricted parameter freedom. More importantly, the model predicts distinctive departures from standard Friedmann evolution, including a present-day jerk parameter significantly different from the flat $\Lambda$CDM value $j_0=1$. These predictions provide direct observational tests of a quantum-cosmological origin of cosmic acceleration.

Our results suggest that late-time cosmic acceleration may arise as a macroscopic quantum effect and illustrate how quantum cosmology can lead to observable consequences at cosmological scales.

Quantum theory treats pre-selection and post-selection symmetrically. Both are described by density matrices: an initial state $\hat{\rho}_0$ at time $t=0$ and a final state $\hat{\rho}_f$ at time $T$. If an observable $\hat{A}=\sum_i a_i\hat{P}_i$ is measured at time $t\in[0,T]$, the probability of outcome $a_i$ is
\begin{equation}
\mathrm{Prob}(a_i,t)=
C\,\mathrm{Tr}\!\left[
e^{-i\hat{H}(T-t)}
\hat{P}_i
e^{-i\hat{H}t}
\hat{\rho}_0
e^{i\hat{H}t}
\hat{P}_i
e^{i\hat{H}(T-t)}
\hat{\rho}_f
\right],
\label{prepost}
\end{equation}
where $\hat{H}$ is the Hamiltonian, $\hat{P}_i$ are the spectral projectors associated with $\hat{A}$, and $C>0$ is a normalization constant \cite{ABL}.

\begin{figure}[!tbp]
 \includegraphics[width=0.9\textwidth]{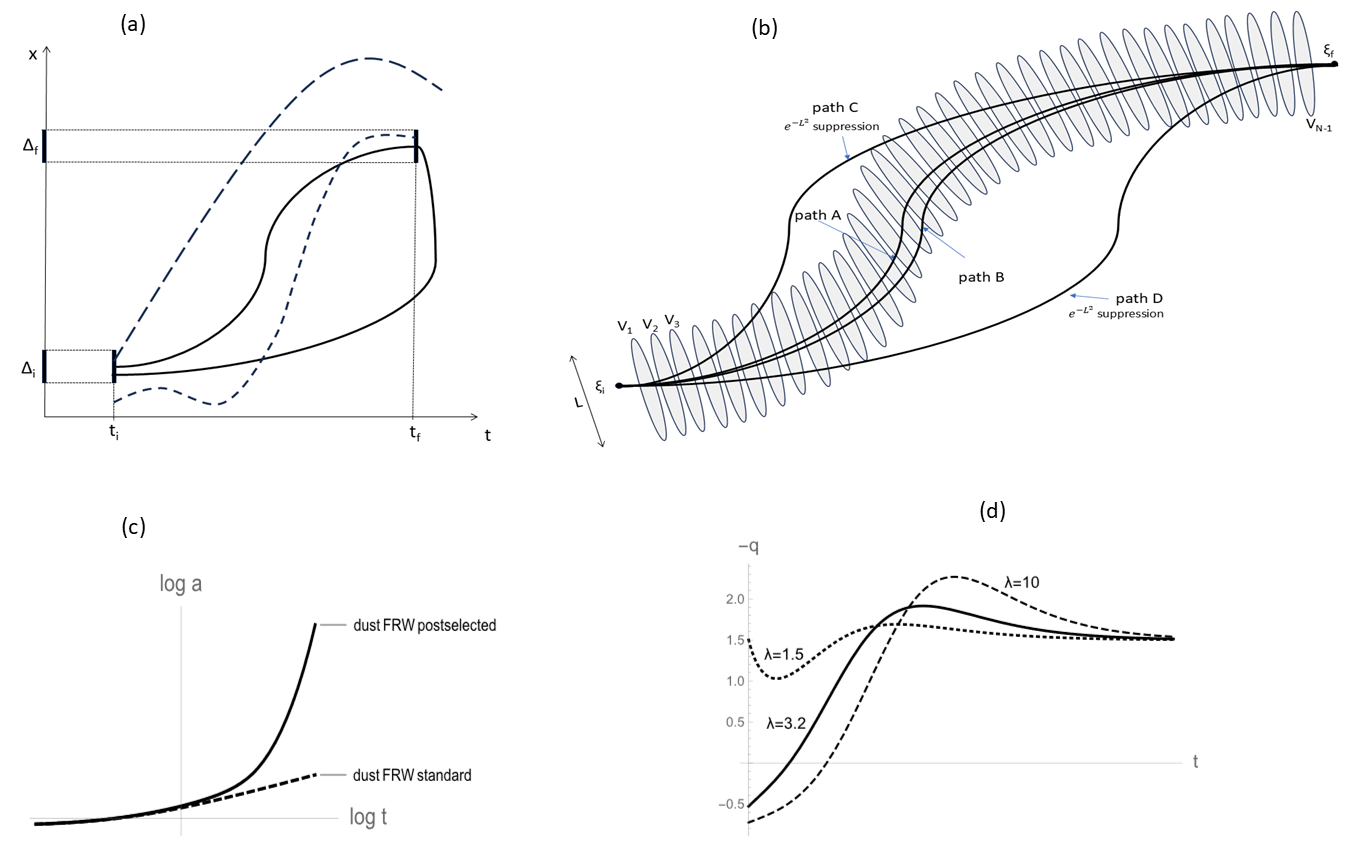}
  \caption{
Post-selected cosmology and emergence of effective dynamics.
(a) Post-selection: only histories with initial conditions in a region $\Delta_i$ at time $t_i$ and final conditions in a region $\Delta_f$ at time $t_f$ contribute to the probabilities.
(b) Coarse-graining: probabilities refer to phase-space tubes of characteristic size $L$. For fixed initial and final conditions, probability concentrates on a single tube, while alternative histories are exponentially suppressed.
(c) Comparison with standard evolution: the dynamics agrees with standard evolution at early times but deviates at late times as post-selection effects become significant.
(d) Cosmic acceleration: the parameter $-q=\ddot{a}a/\dot{a}^2$ as a function of time. For $\lambda\leq1.73$ the expansion accelerates at all times, while for larger values of $\lambda$ the evolution includes an initial decelerating phase.
}
      \label{fig1}
\end{figure}

The study of quantum systems that are both pre- and post-selected is well established both theoretically and experimentally \cite{twostate, SGV}.
In Ref. \cite{Ana25}, it was shown that post-selection combined with coarse-graining generically leads to effective quasiclassical dynamics that depart from standard Hamiltonian evolution. When probabilities are evaluated for coarse-grained observables \cite{Omn89, GeHa2, Ana23}, interference is suppressed and probability concentrates on a narrow class of histories (Fig.~\ref{fig1}). The resulting dynamics is effectively quasiclassical and approximately deterministic  \cite{Ana25}.

Let $\Gamma$ be the classical phase space and $f_t$ the one-parameter group of canonical transformations generated by the Hamiltonian. The effective evolution of a system pre-selected at $\xi_i\in\Gamma$ (time $t = 0$) and post-selected at $\xi_f\in\Gamma$ (time $t = T)$ is
\begin{equation}
\xi(t)=f_t\big(\gamma_{\xi_i,\xi_f'}(t)\big),
\label{traj2}
\end{equation}
where $\xi_f'=f_{-T}(\xi_f)$ and $\gamma_{\xi_i,\xi_f'}(t)$ is a curve connecting the initial and final points, determined by the canonical group of the system---see Refs. \cite{Isham83, IsKa} and also  \cite{ Ana23} for a  modern exposition. A summary derivation is provided in the appendix A.1.

This framework applies to generic quantum dynamical systems. Here, 
we apply it to homogeneous and isotropic FRW cosmology with dust. For zero spatial curvature, standard FRW dynamics reduces to that of a particle moving on a line with Hamiltonian evolution: $x(\tau)=x_i + p_i\tau$. The variable $x$ is related to the scale factor through a transformation
\begin{equation}
a=\left(\frac{3}{2}x\right)^{2/3}.
\end{equation}

Applying the general post-selected quasiclassical framework then leads to an effective solution that depends on exactly two additional parameters that correspond to final conditions. The resulting expansion history is highly constrained and predicts observable departures from standard Friedmann evolution.

Here, we consider the minimal realization, involving a single additional parameter $b$. The effective evolution equations become (see Appendix A.2)
\begin{equation}
x(\tau)=x_i\cosh(b\tau)+p_i\tau.
\label{eveq}
\end{equation}

The corresponding Hubble parameter is
\begin{equation}
\frac{H(z)}{H_0}
=
\frac{
g_{\lambda}\!\left(
\frac{\beta}{(1+z)^{3/2}}
\right)
}{
g_{\lambda}(\beta)
}
(1+z)^{3/2},\label{hhzz}
\end{equation}
with
\begin{equation}
g_{\lambda}(x)=
f_{\lambda}'\!\left(f_{\lambda}^{-1}(x)\right),
\qquad
f_{\lambda}(x)=\cosh x+\lambda x.
\end{equation}
The fitting parameters of the model are:
\begin{itemize}
\item the Hubble constant $H_0$,
\item the dimensionless post-selection parameter $\lambda=\frac{p_i}{x_i b}$, and
\item  the parameter $s_0 = b t_0$, which determines the present epoch $t_0$; in Eq. (\ref{hhzz}), $\beta = f_{\lambda}(s_0)$.
\end{itemize}

For comparison, the $\Lambda$CDM evolution equation for the variable $x$ is
\begin{equation}
x(t)=\frac{1}{q}\sinh\left[
\sinh^{-1}(q x_i)
+\frac{p_i t}{1+q^2x_i^2}
\right],
\label{lcdm}
\end{equation}
where $q$ parametrizes the cosmological constant contribution. Like Eq.~(\ref{eveq}), this solution depends on three integration constants. However, $H(z)$ is obtained from the first integral of the motion, and it is invariant under time translation $x(t) \rightarrow x(t+a)$. Therefore, it depends only on two parameters; the usual choice is $H_0$
 and the matter density parameter $\Omega_m$. 
 
 In contrast, POQCO does not reduce to a first-order Friedmann constraint, and the observable Hubble function retains an explicit dependence on all three solution parameters. This structure implies that the cosmic age becomes an independent fitting  parameter through $s_0$, and that there is no quantity analogous to $\Omega_m$.

POQCO also admits an analytic solution when radiation is included. In the early-time limit, the model reproduces the standard radiation- and matter-dominated scalings,
\begin{equation}
a(t)\sim t^{1/2},
\qquad
a(t)\sim t^{2/3},
\end{equation}
so that the Hubble parameter approaches the standard Friedmann form at high redshift. Early-Universe observables, such as the comoving sound horizon, therefore remain close to their standard values.

\begin{figure}[ht]
    \centering
    \includegraphics[width=0.7\textwidth]{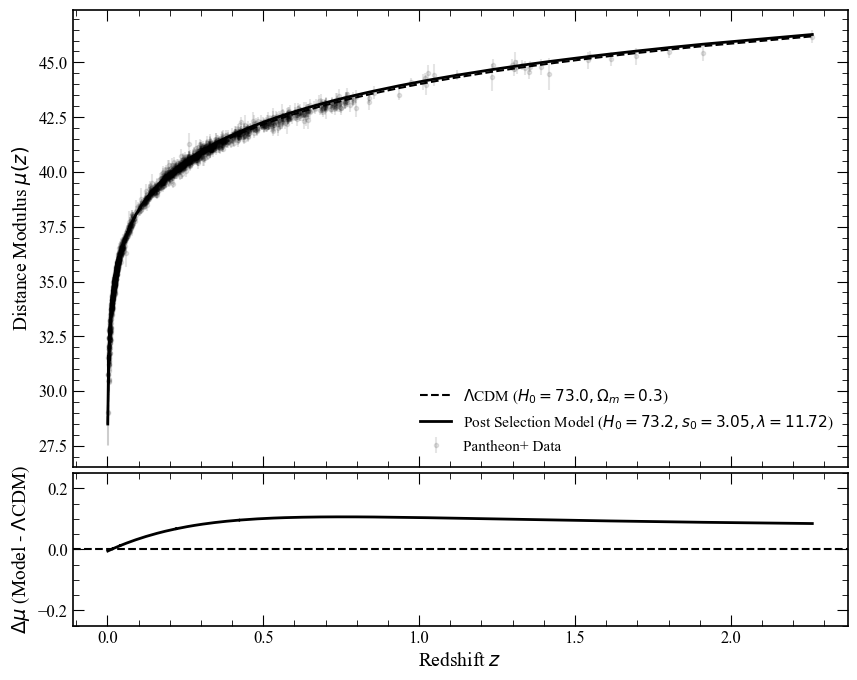}
    \caption{Hubble diagram for the Pantheon+ Type Ia Supernovae dataset (top) and the corresponding residuals relative to the baseline $\Lambda$CDM model (bottom).}
    \label{fig:hubble_residuals}
\end{figure}

We first test POQCO against the Pantheon+ type Ia supernova compilation \cite{Scolnic22}. The best-fit parameters are
$$ 
H_0 = 72.32^{+0.25}_{-0.25}\,
{\rm km\,s^{-1}\,Mpc^{-1}},
\qquad
s_0 = 3.11^{+0.16}_{-0.17},
\qquad
\lambda = 3.2^{+1.1}_{-0.9}.
$$
Figure \ref{fig:hubble_residuals} shows that the model reproduces the observed distance-redshift relation with residuals comparable to those of $\Lambda$CDM. In fact, according to the Akaike information criterion (AIC) \cite{Akaike74}, POQCO moderately outperforms $\Lambda$CDM, as $\Delta AIC = -2.5$. However,   the Bayes information criterion (BIC) \cite{Schwarz78} leads to the opposite conclusion, since $\Delta BIC = 3$. The reason for this difference is that BIC imposes a more rigorous penalty to the additional parameter $s_0$ \cite{Liddle07}. For details, see   Appendix D.2.

\begin{figure}[ht]
    \centering
    \includegraphics[width=0.6\textwidth]{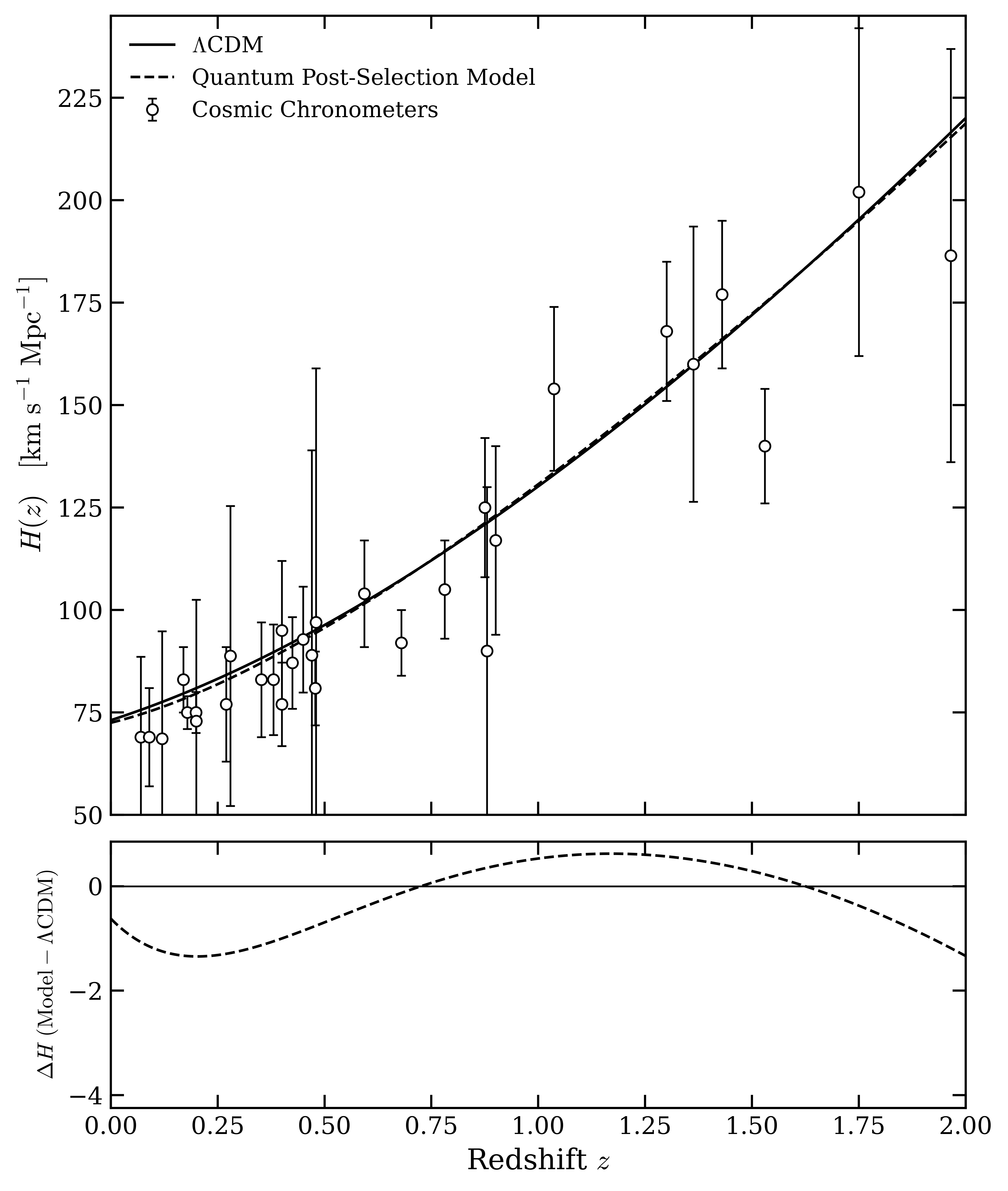} 
    \caption{Evolution of the Hubble parameter $H(z)$ as predicted by the post-selection quantum model compared to 31 Cosmic Chronometer data points. The lower panel displays the residuals relative to the standard $\Lambda$CDM model.}
    \label{fig:hz_fit}
\end{figure}

We also test POQCO against a compilation of cosmic-chronometer measurements of $H(z)$ \cite{Yu18}. Figure~\ref{fig:hz_fit} compares the best-fit expansion history with the data. We obtain
$$ 
H_0 = 68.9^{+3.9}_{-3.7}
\,{\rm km\,s^{-1}\,Mpc^{-1}},
 \qquad
s_0 = 2.11^{+0.35}_{-0.44},
\qquad
\lambda = 4.78^{+1.53}_{-1.57}.
$$ 
The resulting fit is statistically competitive with $\Lambda$CDM and yields parameter values consistent with those inferred from the supernova sample within the larger uncertainties of the cosmic-chronometer data.

Thus, according to POQCO, accelerated expansion emerges from the effective dynamics induced by the well understood mechanisms of  quantum post-selection and coarse-graining without the need to introduce a cosmological constant or new physics. Related ideas have also recently  been explored in a different framework that employs   a Chern--Simons soliton for implementing the final condition \cite{DaMa26}.

A distinctive prediction of POQCO is an earlier onset of accelerated expansion than in standard cosmology. Using the Pantheon+ best-fit parameters, we find
$$ 
z_{\rm acc}=1.9\pm0.8,
$$ 
substantially larger than the value $z_{acc} \sim 0.7$ predicted by the concordance $\Lambda$CDM model. However, unlike $\Lambda$CDM, this transition is not associated with a special epoch at which two independent energy densities become comparable; hence, the coincidence problem \cite{ZWS} does not arise in POQCO.

An additional observational discriminator is provided by the cosmographic jerk parameter,
$$ 
j=\frac{\dddot a}{aH^3},
$$ 
which characterizes departures from standard Friedmann evolution \cite{Visser}. In flat $\Lambda$CDM cosmology, neglecting radiation, one has the exact relation $j=1$, independent of the values of $H_0$ and $\Omega_m$. In contrast, POQCO predicts a nontrivial jerk evolution. For the Pantheon+ best-fit parameters, we obtain
$$ 
j_0 = 2.2 \pm 0.3,
$$ 
which differs from the flat $\Lambda$CDM prediction at approximately the $4\sigma$ level. This difference suggests that future cosmographic observations, particularly improved cosmic-chronometer measurements of $H(z)$, may provide a direct observational test of the quantum-cosmological origin of acceleration. 

Note that canonical quintessence models and non-phantom barotropic fluids typically predict $j_0<1$, while generalized Chaplygin-gas models generally yield  $j_0 < 1.7$ for parameter ranges compatible with current observations. Therefore, the POQCO prediction for $j_0$ lies in a distinct region of the cosmographic parameter space. For comparison, a constant-$w$ phantom fluid would require $w\simeq -1.3$ to produce a similar value of $j_0$.

We have shown that POQCO provides a viable late-time cosmology derived from a previously established framework for post-selected quasiclassical dynamics. Despite its non-Friedmann structure and highly constrained parameter space, the model yields observational fits competitive with $\Lambda$CDM without introducing a cosmological constant, dark energy, or modifications of general relativity.

Unlike phenomenological dark-energy reconstructions, where the expansion history can often be adjusted through an appropriate choice of model parameters, the POQCO background evolution follows from a highly constrained  dynamical framework. The resulting modification of $H(z)$ is therefore not freely reconstructed, but determined by a small number of parameters associated with the effective post-selected dynamics.
A distinctive consequence of this dynamics is the  lack of a conserved first integral of the motion, encoded in the Friedman equation. This results to the strong divergence of quantities such as the jerk,  not only from the    $\Lambda$CDM prediction but from typical predictions of dark energy models that employ  a form of  the Friedman equation.

Our results suggest that cosmic acceleration may arise as a macroscopic quantum effect associated with post-selection. Furthermore, they illustrate how quantum cosmology can lead to observable consequences on cosmological scales and at late times, and that its implications are not restricted to physics near the Big Bang. More generally, they provide a concrete example of how final conditions in physics can give rise to distinctive and observationally testable predictions.

 \section*{Acknowldgements}

 This work was supported by computational time granted by the National Infrastructures for Research and Technology S.A. (GRNET S.A.) in the National HPC facility - ARIS - under project ID
pr017008/simnstar2.  C. A. acknowledges support from the COST Action CA23115 ``Relativistic Quantum Information".

\begin{appendix}
\numberwithin{equation}{section}

\section{Derivation of the effective cosmological dynamics}

This section provides the derivation of the effective post-selected dynamics used in the main text.

\subsection{Classical limit of a post-selected system and coarse-grained histories}

Here we summarize the derivation of Eq.~(2) of the main text. The full derivation is given in Ref.~\cite{Ana25}.

Quantum measurements are typically described in terms of \emph{pre-selected} ensembles, whose preparation is encoded in an initial density matrix. However, quantum theory also admits \emph{post-selected} ensembles \cite{ABL, twostate}, in which outcomes are retained only if a final condition is satisfied. Pre- and post-selection enter symmetrically: the system is described by an initial density matrix $\hat{\rho}_0$ at time $t=0$ and a final density matrix $\hat{\rho}_f$ at time $T$.

If an observable $\hat{A} = \sum_i a_i \hat{P}_i$ is measured at time $t \in [0,T]$, the probability of outcome $a_i$ is
\bey
\mbox{Prob}(a_i, t) = C \mbox{Tr} \left[ e^{-i\hat{H}(T - t)}\hat{P}_ie^{-i\hat{H}t} \hat{\rho}_0e^{i\hat{H}t} \hat{P}_ie^{i\hat{H}(T - t)} \hat{\rho}_f\right],
\label{prepost_SI}
\eey
where $\hat{H}$ is the Hamiltonian and $C$ is a normalization constant.

To describe effective dynamics, we partition the time interval $[0,T]$ into $N$ steps and consider $N-1$ successive coarse-grained measurements at times $t_n = nT/N$, associated to regions $V_1, \ldots, V_{N-1}$ in the classical state space. The probability for a sequence of outcomes is
\bey
\mbox{Prob}(V_1,t_1;\ldots;V_{N-1},t_{N-1})
=
K\,|{\cal A}(V_1,\ldots,V_{N-1})|^2,
\eey
where $K$ is a normalization constant and the amplitude is
\bey
{\cal A}
=
\langle \psi_f|e^{-i\hat{H}T}
\hat{\Pi}_{V_{N-1}}(t_{N-1})
\cdots
\hat{\Pi}_{V_1}(t_1)
|\psi_i\rangle.,
\eey
where $\hat{\Pi}_V(t) = e^{i\hat{H}t} \hat{\Pi}_V e^{-i\hat{H}t}$. 
For sufficiently coarse-grained observables, approximate determinism holds \cite{Omn89, Omn1}, so that $\hat{\Pi}_{V}(t) \simeq \hat{\Pi}_{f_{-t}(V)}$, where $f_t$ is the time evolution (canonical transformation) generated by the corresponding classical Hamiltonian.
\bey
{\cal A}
\simeq
\langle \psi_f'|
\hat{\Pi}_{f_{-t_{N-1}}(V_{N-1})}
\cdots
\hat{\Pi}_{f_{-t_1}(V_1)}
|\psi_i\rangle,
\eey
where $|\psi_f'\rangle = e^{i\hat{H}T}|\psi_f\rangle$.

We assume that the initial and final states are localized around phase-space points $\xi_i$ and $\xi_f'$, respectively. In this case, the probability distribution over histories is sharply peaked around a single trajectory determined by these boundary conditions. This leads, in the continuum limit, to an effective deterministic evolution law for the system, as given in Eq.~(2) of the main text \cite{Ana25}.

\subsection{ Cosmological dynamics}

We now specialize the post-selected effective dynamics to a homogeneous and isotropic dust cosmology. The Friedmann equation for a dust-filled FRW universe can be written as
\bey
\dot a^2+\kappa=\frac{\mu}{6a},
\eey
where $\kappa$ is the spatial curvature parameter and $\mu$ is a constant proportional to the conserved dust energy.

It is useful to introduce the variable
\bey
x=\left(\frac{2}{3}a\right)^{3/2}.
\eey
In terms of $x$, the Friedmann equation becomes
\bey
\dot x^2+\kappa x^{2/3}=\epsilon,
\eey
where $\epsilon$ is a constant. For the spatially flat case, $\kappa=0$, this reduces to the equation of motion of a free particle on the half-line $x>0$.

Since the physical configuration space is the half-line ($x > 0$), the appropriate canonical structure is not the ordinary Weyl algebra on the full line, but the affine algebra
\bey
[\hat x,\hat\pi]=i\hat x .
\eey

The action of the associated Lie group on the classical state space defines the trajectories $\gamma_{\xi_i, \xi_f}(t)$ in Eq. (2) of the main text. Explicitly, they read
\bey
x(\tau)=x_i\cosh(b_1\tau),
\qquad
p(\tau)
=
p_i+
\frac{p_f-p_i}{\tanh(b_1T)}
\tanh(b_1\tau),
\eey
where
\bey
b_1=T^{-1}\cosh^{-1}\left(\frac{x_f}{x_i}\right).
\eey

Substituting these expressions  into the general post-selected trajectory formula---Eq. (2) of the main text---gives
\bey
x(\tau)
=
x_i\cosh(b\tau)
+
\left[
p_i+
\frac{p_f-p_i}{\tanh(bT)}
\tanh(b\tau)
\right]\tau ,
\label{eq:post_selected_x_general}
\eey
and
\bey
p(\tau)
=
p_i+
\frac{p_f-p_i}{\tanh(bT)}
\tanh(b\tau),
\label{eq:post_selected_p_general}
\eey
with
\bey
b
=
T^{-1}\cosh^{-1}
\left(
\frac{x_f-p_fT}{x_i}
\right).
\eey
Thus the post-selected cosmological evolution is fixed by the initial data together with the final boundary condition.

Equivalently, the effective solution may be written in the form
\bey
x(\tau)
=
x_i\cosh(b\tau)
+
\left[
p_i+\sigma\tanh(b\tau)
\right]\tau ,
\label{eq:post_selected_x_sigma}
\eey
where the two parameters $b$ and $\sigma$ encode the dependence on the final condition. The parameter $b$ controls the late-time deviation from the standard dust solution and is responsible for the accelerated expansion. In this sense, $b$ plays a role analogous to a cosmological constant at late times, although no cosmological constant or additional dark-energy component has been introduced.

In the minimal model analysed in the main text we restrict to the one-parameter subfamily $\sigma=0$,
so that
\bey
x(\tau)=x_i\cosh(b\tau)+p_i\tau .
\label{eq:minimal_post_selected_solution}
\eey

\subsection{Expression for $H(z)$ for $\kappa = 0$}

The Hubble factor for Eq. (\ref{eq:minimal_post_selected_solution}) is
\bey
H(s) = \frac{2}{3}b \frac{\sinh s + \lambda}{\cosh s + \lambda s},
\eey
 where $s = b\tau$ and $\lambda = \frac{p_i}{x_i b}$. 

Denote $f_{\lambda}(s) = \cosh s + \lambda s$. Let $s = s_0$ correspond to the current cosmological time, and write $x(s_0) = x_0$ and $H(s_0) = H_0.$

We have
\bey
\frac{x}{x_0} = \frac{f_{\lambda}(s)}{f_{\lambda}(s_0)} =\frac{1}{(1+z)^{3/2}},
\eey
in terms of the redshift $z$.
This leads to
\bey
\frac{H(z)}{H_0} = \frac{g_{\lambda}\left( \frac{\beta}{(1+z)^{3/2}} \right)}{g_{\lambda}(\beta)}(1+z)^{3/2},
\eey
where $\beta = f_{\lambda}(s_0)$ and 
\bey
g_{\lambda}(x) = f_{\lambda}'(f_{\lambda}^{-1}(x)) .
\eey
The model has two parameters, $\lambda$ and  $\beta$ (or $\lambda$ and $s_0$), in addition to $H_0$.

\subsection{High-redshift limit and radiation}

Next, we show that the model reduces to standard FRW cosmology with radiation and matter at high-redshift.

Radiation is neglected in the main analysis because the datasets considered are late-time probe, but it is naturally incorporated. For $\kappa = 0$, and using the variable $x$, the FRW equation becomes
\bey
\dot{x}^2 = \epsilon + \gamma/x^{2/3},
\eey
 where $\gamma \sim \Omega_{rad}/\Omega_m$. The classical solution is given by
 \bey
 x(t) = \eta^{-1}_{\gamma}\left[\eta_{\gamma}(x_i) + \frac{p_i t}{\sqrt{1+\gamma/x_i^{2/3}}}\right], \label{early}
 \eey
 where $\eta_{\gamma}(x) = x(1-2\gamma/x^{2/3})\sqrt{1+\gamma/x^{2/3}}$.
The postselected solution with $\sigma = 0$ is 
  \bey
 x(t) = \eta^{-1}_{\gamma}\left[\eta(x_i \cosh(bt)) + \frac{p_i t}{\sqrt{1+\frac{\gamma}{x_i^{2/3}\cosh^{2/3}(bt)}}}\right]. \label{earlyp}
 \eey
It is straightforward to see that at early times, $b t << 1$, the quantum post-selection model yields the standard FRW evolution (\ref{early}), and for $\gamma = 0$, we obtain the pure-dust model analyzed here.
 
\section{Supernova Likelihood}
First, we test POQCO versus Supernovae Type Ia. We use the Pantheon+ dataset, which comprises 1701 light curves from 1550 unique, spectroscopically confirmed Type Ia supernovae (SNe Ia).  These catalogues relate observable values for the distance modulus $\mu_{obs}$ at the the redshift of the event ranging from a z of $10^{-3}$ to $2.27$\cite{Scolnic22}. We evaluated the parameter space of POCQO by performing a Markov Chain Monte Carlo (MCMC) analysis using the open-source package \texttt{emcee} \cite{EMCEE}.

The observed distance modulus is $\mu_{\text{obs}}(z) = m_B(z) - M$, where $m_B$ is the apparent magnitude and $M$ is the intrinsic absolute magnitude. The theoretical expectation is:
\begin{equation}
    \mu_{\text{th}}(z; \mathbf{\theta}) = 5 \log_{10} \left( \frac{d_L(z; \mathbf{\theta})}{\text{Mpc}} \right) + 25.
\end{equation}
For a general FRW geometry containing spatial curvature $\Omega_k$, the luminosity distance $d_L$ is:
\begin{equation}
    d_L(z) = \frac{c(1+z)}{H_0 \sqrt{|\Omega_k|}} \text{sinn} \left( \sqrt{|\Omega_k|} \int_0^z \frac{H_0 dz'}{H(z'; s_0, \lambda)} \right),
\end{equation}
where $\text{sinn}(x)$ denotes $\sinh(x)$, $\sin(x)$, or $x$ for open, closed, or flat geometries, respectively. 

Because $M$ and $H_0$ are completely degenerate, we analytically marginalize over $M$. The modified $\chi^2$ statistic is:
\begin{equation}
    \chi^2_{\text{SNe}} = a - \frac{b^2}{d} + \log\left(\frac{d}{2\pi}\right),
\end{equation}
with $a = \Delta\boldsymbol{\mu}^T \mathbf{C}^{-1} \Delta\boldsymbol{\mu}$, $b = \Delta\boldsymbol{\mu}^T \mathbf{C}^{-1} \mathbf{1}$, and $d = \mathbf{1}^T \mathbf{C}^{-1} \mathbf{1}$, where $\Delta\boldsymbol{\mu} = \mu_{\text{obs}} - \mu_{\text{th}}$ and $\mathbf{C}$ is the total covariance matrix. Using a $z > 0.01$ cutoff to mitigate peculiar velocities, our SNIa-only analysis tightly constrains the functional shape of the expansion, yielding $s_0 = \text{3.11} \pm \text{0.17}$ and $\lambda = \text{3.21} \pm \text{1.12}$. However, due to the marginalization, $H_0$ remains unconstrained.

\begin{figure}[ht]
    \centering
    \includegraphics[width=0.7\textwidth]{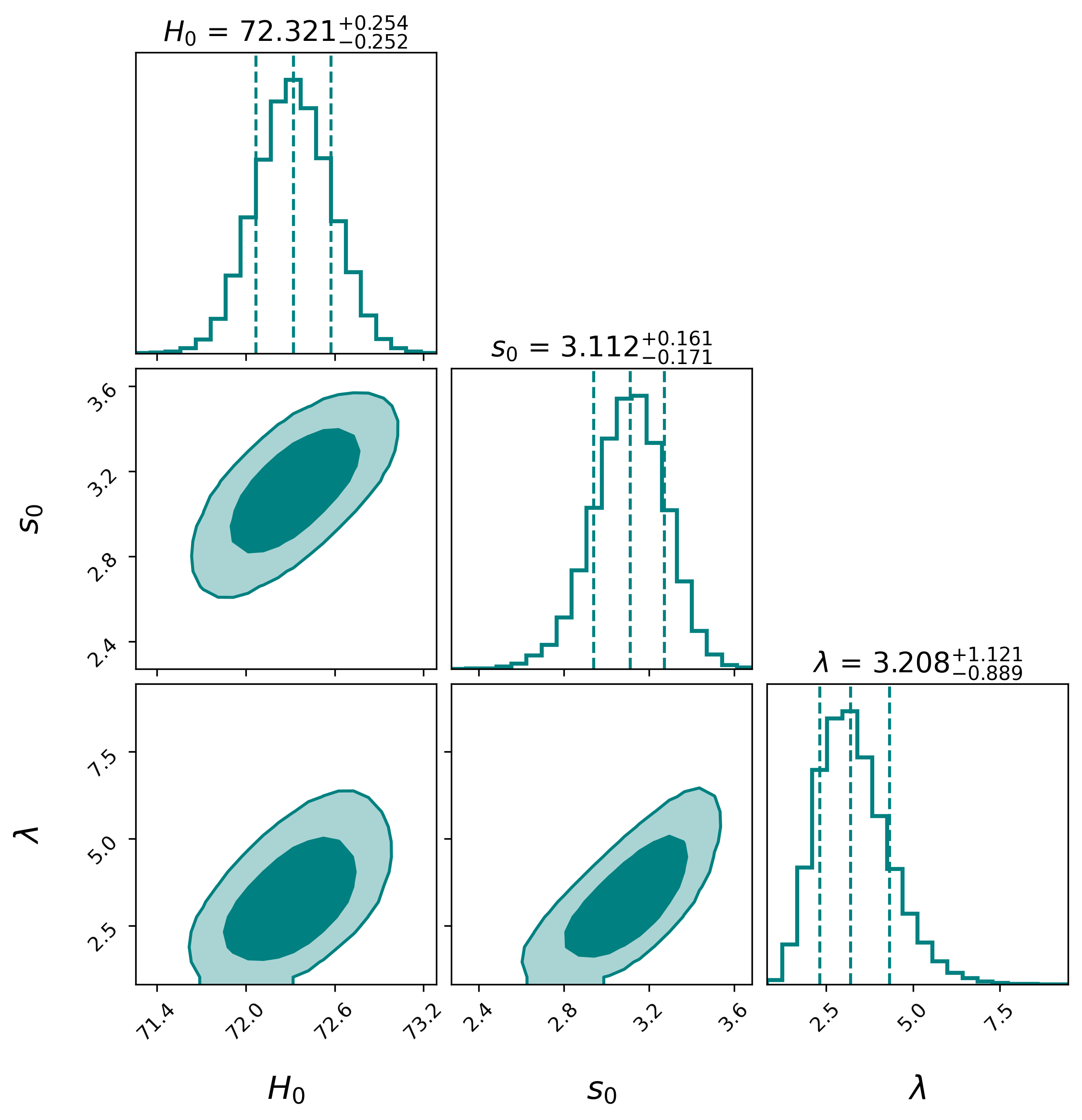}
    \caption{Posterior distributions and parameter correlations derived from the Pantheon+ SNIa dataset.}
    \label{fig:Contour Plot}
\end{figure}


\section{Cosmic Chronometer Analysis}
 The Cosmic Chronometer (CC) dataset serves as a vital model-independent probe for our post-selection framework. Unlike distance-based observations, CC measurements determine the expansion rate $H(z)$ directly through the differential aging of passively evolving galaxies, making them independent of any prior cosmological assumptions or distance ladder calibrations.
 
  We use a compilation of $N=31$ measurements spanning $0.07 \leq z \leq 1.965$. The likelihood is defined by:
\begin{equation}
    \chi^2_{\text{CC}}(H_0, s_0, \lambda) = \sum_{i=1}^{31} \left( \frac{H_{\text{obs}}(z_i) - H_{\text{PS}}(z_i; H_0, s_0, \lambda)}{\sigma_{H_i}} \right)^2.
\end{equation}

Our CC-only analysis acts as an absolute anchor for the model, yielding $H_0 \approx 68.9 \pm 1.7$ km s$^{-1}$ Mpc$^{-1}$, which aligns well with early-universe constraints. However, because CC data tracks the expansion slope, $H_0$ is highly degenerate with the post-selection parameters. Consequently, the isolated CC constraints on $s_0 = \text{2.11} \pm \text{0.44}$ and $\lambda = \text{4.8} \pm \text{1.6}$ remain broad.

\begin{figure}[ht]
    \centering
    \includegraphics[width=0.7\textwidth]{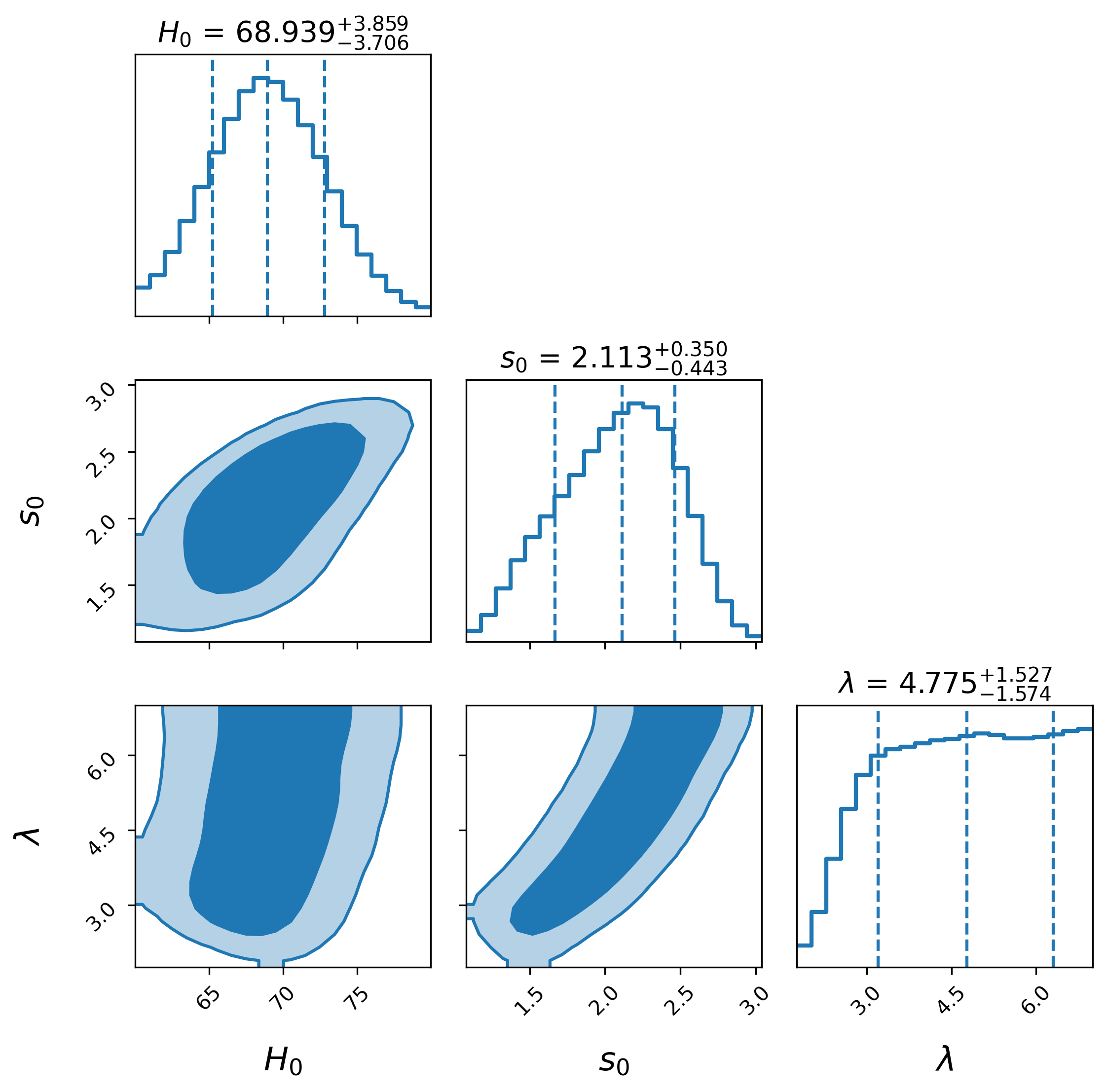}
    \caption{Posterior distributions derived from the CC dataset, highlighting the degeneracy between $H_0$ and the quantum parameters.}
    \label{fig:cc_contours}
\end{figure}

\section{Joint Statistical Analysis}

\subsection{Results}

To further constrain our model's parameter space and to break the inherent degeneracies found in individual datasets, we define a total likelihood function. Assuming that the Pantheon+ SNIa sample and the Cosmic Chronometer (CC) measurements are statistically independent, the joint likelihood is defined as the product of the individual likelihoods:

\begin{equation}
\mathcal{L}_{\mathrm{tot}}(\theta) = \prod_{p=1}^{P} \exp\left(-\frac{1}{2}\chi^2_p\right)
\end{equation}
where $\theta = \{H_0, s_0, \lambda\}$ represents the vector of free parameters. This corresponds to the summation of the individual $\chi^2$ statistics:

\begin{equation}
\chi^2_{\mathrm{tot}} = \chi^2_{\mathrm{SNIa}} + \chi^2_{\mathrm{CC}}
\end{equation}

For the Supernovae component, we utilize the marginalized $\chi^2$ to account for the nuisance parameter of the absolute magnitude $M$, while the CC likelihood is computed directly using the observed $H(z)$ values. We explore the resulting multidimensional posterior distribution. We deployed 64 walkers for 10,000 steps, ensuring convergence by monitoring the integrated autocorrelation time. This joint approach allows the geometric constraints of the supernovae to complement the direct expansion rate measurements of the chronometers, effectively narrowing the allowed ranges for the quantum post-selection parameters $s_0$ and $\lambda$.

The joint analysis shifts the preferred value of $H_0$ toward the CC determination: we obtain $H_0 = 67.1 \pm 1.7$ This reflects the different roles of the two probes: the Pantheon+ sample primarily constrains the shape of the expansion history, while CC provide an absolute calibration of $H(z)$. The combined likelihood therefore favors a compromise expansion history that preserves the supernova fit while adjusting the overall normalization.



\subsection{Information Criteria and Parameter Counting}

To assess the statistical performance of POQCO relative to the baseline $\Lambda$CDM framework, we employ the Akaike Information Criterion (AIC) \cite{Akaike74} and the Bayesian Information Criterion (BIC) \cite{Schwarz78}. These metrics are defined as:
\begin{equation}
    \mathrm{AIC} = 2k - 2\ln\mathcal{L}_{\text{max}}, \quad \mathrm{BIC} = k\ln N - 2\ln\mathcal{L}_{\text{max}},
\end{equation}
where $k$ denotes the number of free parameters and $N$ is the total number of data points ($N = 1732$ for the joint SNIa + CC sample). While both criteria penalize the inclusion of additional parameters to prevent overfitting, the BIC imposes a more rigorous penalty proportional to the logarithm of the sample size \cite{Liddle07}.

The results of our comparative analysis are summarized in Table~\ref{tab:results_comparison}. For the joint analysis, the post-selection model ($k=3$) yields $\mathrm{AIC} = 1771.48$ and $\mathrm{BIC} = 1787.85$, while the $\Lambda$CDM model ($k=2$) results in $\mathrm{AIC} = 1771.42$ and $\mathrm{BIC} = 1782.34$. The negligible difference in AIC ($\Delta \mathrm{AIC} \approx 0.06$) indicates that both models provide an essentially equivalent description of the data from an information-theoretic perspective. However, the $\Delta \mathrm{BIC} \approx 5.51$ suggests ``moderate'' evidence in favor of the simpler $\Lambda$CDM model according to the Jeffreys scale \cite{Kass95}.



\begin{table}[H]
\centering
\caption{Observational constraints, maximum log-likelihood ($\ln\mathcal{L}_{\text{max}}$), and information criteria AIC and BIC for the post selection model and the baseline $\Lambda$CDM model.}
\label{tab:results_comparison}
\begin{tabular*}{\textwidth}{@{\extracolsep{\fill}}lcccc}
\hline
\hline
\textbf{Method} & \boldmath{$H_0$ \textbf{[km s$^{-1}$ Mpc$^{-1}$]}} & \boldmath{$\ln \mathcal{L}_{\text{max}}$} & \textbf{AIC} & \textbf{BIC} \\ 
\hline
\textit{Post-Selection Model} & & & & \\
SNIa (Pantheon+)     & $72.32^{+0.25}_{-0.25}$ & $-875.28$ & $1756.57$ & $1772.88$ \\
Cosmic Chronometers  & $68.9^{+3.9}_{-3.7}$ & $-7.13$  & $20.27$  & $24.57$  \\
Joint (SNIa + CC)    & $67.1 \pm 1.7$      & $-882.74$ & $1771.48$ & $1787.85$ \\ 
\hline
\textit{$\Lambda$CDM Model} & & & & \\
SNIa (Pantheon+)     & $72.84 \pm 0.25$        & $-877.51$ & $1759.02$ & $1769.90$ \\
Cosmic Chronometers  & $67.7 \pm 3.1$        & $-7.25$  & $18.50$  & $21.36$  \\
Joint (SNIa + CC)    & $66.5 \pm 1.7$        & $-883.71$ & $1771.42 $ & $1782.34$ \\
\hline
\hline
\multicolumn{5}{l}{}
\end{tabular*}
\end{table}


\subsection{Interpretation of the joint-fit shift in $H_0$}

The shift in the marginalized $H_0$ posterior in the joint analysis arises from the fact that both datasets impose distinct geometric and dynamical constraints. Within our framework, this shift illustrates the interplay between the absolute normalization of the expansion rate and its relative gradient. The Cosmic Chronometer (CC) dataset provides the necessary absolute expansion scale but suffers from a broad $H_0 - \{s_0, \lambda\}$ degeneracy manifold. In this "banana" distribution, various combinations of a high $H_0$ and specific values of the free parameters can statistically satisfy the $H(z)$ data.

In contrast, the Pantheon+ dataset—analytically marginalized over the absolute magnitude $M$—tightly constrains the relative expansion "shape" $E(z) = H(z)/H_0$ with high precision, while remaining insensitive to the absolute normalization. When the likelihoods are combined, the precise SNIa-derived gradient effectively "slices" through the broader CC degeneracy manifold. The resulting shift in $H_0$ represents the specific absolute height required to align the absolute CC data points with the high-precision slope dictated by the supernovae. This joint estimate gravitates toward the region of parameter space where the absolute scale and the relative expansion history are mutually consistent, providing a more robust determination of the local expansion rate.

\section{Cosmographic and phenomenological diagnostics}

In this section, we summarize several phenomenological quantities associated with the best-fit POQCO background evolution. These quantities are intended as illustrative diagnostics of the post-selected dynamics rather than as independent observational constraints.
The quantities are evaluated using the Pantheon+ best-fit parameters
\[
s_0 = 3.11^{+0.16}_{-0.17},
\qquad
\lambda = 3.2^{+1.1}_{-0.9}.
\]

\begin{table}[H]
\centering

\label{tab:diagnostics}
\begin{tabular}{lcc}
\hline
\hline
Quantity & POQCO & flat $\Lambda$CDM \\
\hline
Present deceleration parameter $q_0$
& $ -1.21 \pm 0.04$
& $\sim -0.55$
\\[2mm]

Present jerk parameter $j_0$
& $ 2.2 \pm 0.3$
& $1$
\\[2mm]

Acceleration transition redshift $z_{\rm acc}$
& $ 1.9\pm 0.8$
& $\sim 0.7$
\\[2mm]

Effective equation of state parameter $w_{\rm eff,0}$
& $-1.14\pm 0.03$
& $-1$
\\[2mm]

Effective sound-speed parameter  $\kappa_0$
& $-1.9 \pm 0.6$
& $-1$  
\\[2mm]

\hline
\hline
\end{tabular}
\caption{Representative cosmographic and phenomenological quantities for the best-fit POQCO background compared to flat $\Lambda$CDM with $\Omega_m=0.3$.}
\end{table}


If the POQCO expansion history is reinterpreted within an effective Friedmann-fluid description, the resulting equation-of-state parameter $w$ can fall below $-1$. This should not be interpreted as a physical phantom fluid, since POQCO contains no dark-energy component. Rather, it is a cosmographical diagnostic that quantifies the departure   from standard Friedmann evolution. 
Another such diagnostic is the 
 effective sound-speed parameter, $\kappa_0:= (dP/d\rho)_{0} = \frac{1}{3} (j_0-1)/(1+q_0)$ \cite{Visser}, which  diverges from the $\Lambda$CDM value $-1$.

\section{Analysis for non-zero curvature}

Analytical solutions within the POQCO framework can also be derived for non-zero spatial curvature. The curvature parameter $\kappa = \pm 1, 0$ acts as a discrete geometric indicator, where $\kappa=1$ signifies a closed universe and $\kappa=-1$ denotes an open universe. Setting the scale factor to $a = \frac{3}{2} x^{2/3}$, the modified Friedmann equation takes the form
\begin{equation}
\dot{x}^2 + \kappa x^{2/3} = \epsilon,
\end{equation}
where $\epsilon = p_i^2 + \kappa x_i^{2/3}$ is a constant. Under this setup, the post-selected trajectory evolves according to
\begin{equation}
x(t) = \epsilon_t^{3/2} u^{-1}\left(\frac{t}{\epsilon_t} + u(x_i \cosh(bt)/\epsilon_t^{3/2})\right), 
\end{equation}
where $\epsilon_t = p_i^2 + \kappa x_i^{2/3} \cosh^{2/3}(bt)$, and the geometric structural function $u(x)$ is defined analytically as
\begin{equation}
u(x) = \int \frac{dx}{\sqrt{1-\kappa x^{2/3}}} = \left\{\begin{array}{cc}    -\frac{3}{2}[x^{1/3} \sqrt{1-x^{2/3}} - \arcsin(x^{1/3})],& \kappa = 1 \\
x, &\kappa = 0 \\\frac{3}{2}[x^{1/3} \sqrt{1+x^{2/3}} - \operatorname{arcsinh}(x^{1/3})],&\kappa = -1\end{array}\right.
\end{equation}

By tracking the dimensionless parameter $s = bt$ via an alternative parameterization defined by $v = \cosh s$, we introduce the dimensionless scale parameters $\alpha = p_i^2 / x_i^{2/3}$ and $\beta = b x_i^{2/3}$. This maps the evolution onto the auxiliary tracking function $w(\alpha, \beta, v)$:
\begin{equation}
w(\alpha, \beta, v) = (\alpha + v^{2/3})^{3/2} u^{-1}\left(\frac{\cosh^{-1}(v)}{\beta (\alpha + v^{2/3})}+ u\left(\frac{v}{(\alpha + v^{2/3})^{3/2} }\right)\right).
\end{equation}
Consequently, the full normalized Hubble expansion rate is dictated by a 4-parameter space ($\alpha, \beta, H_0, v_0$) and is explicitly evaluated via:
\begin{equation}
\label{eq:H_curvature}
\frac{H(z)}{H_0} = \sqrt{\frac{\alpha + v(z)^{2/3} - \kappa w(v_0)^{2/3}/(1+z)}{\alpha + v_0^{2/3} - \kappa w(v_0)^{2/3}}}(1+z)^{3/2},
\end{equation} 
where the redshift dependence of the coordinate variable is given implicitly by the inversion relation $v(z) = w^{-1}\left(w(v_0)/(1+z)^{3/2}\right)$.

\begin{figure}[ht!]
    \centering
    \includegraphics[width=0.7\textwidth]{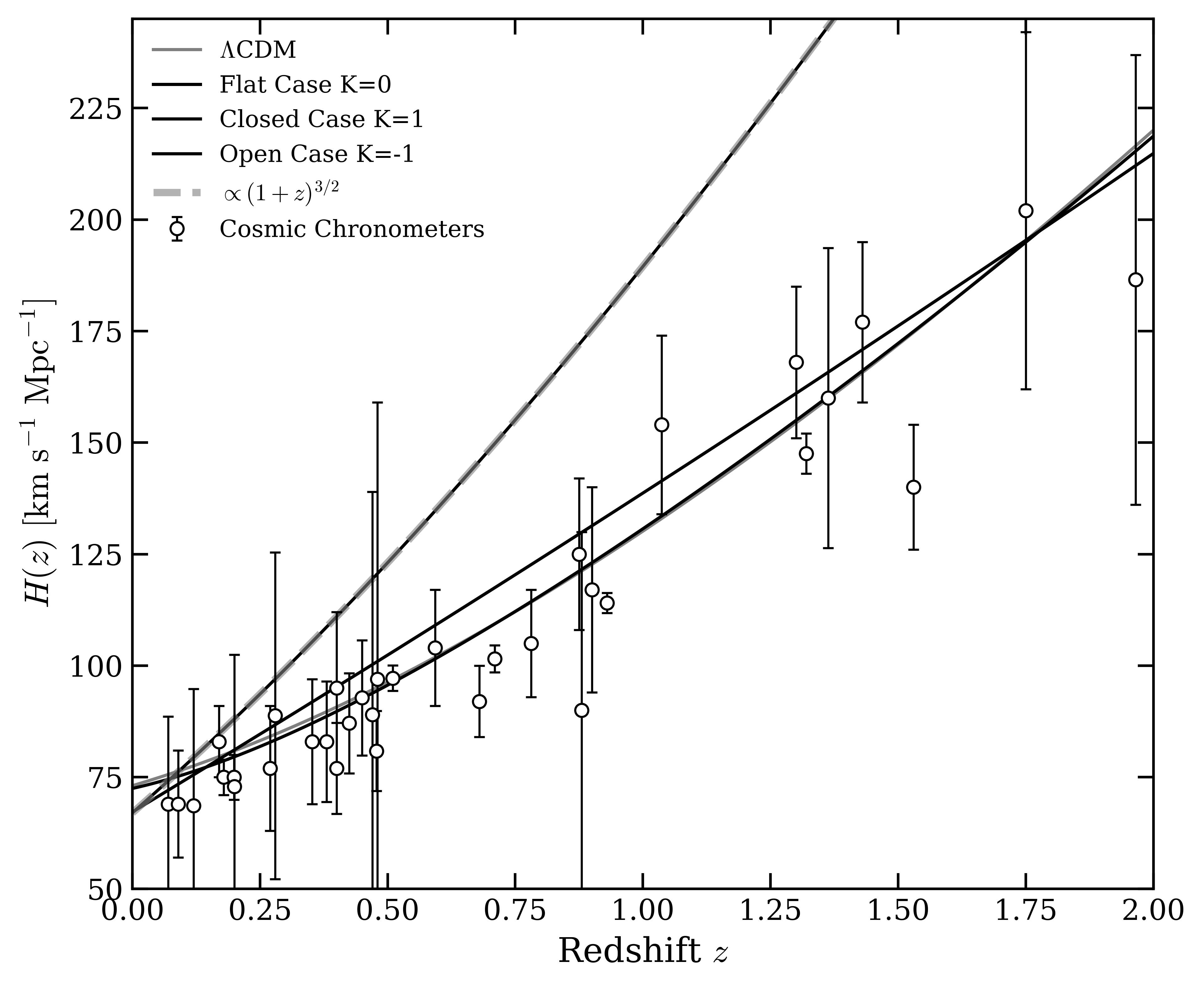}
    \caption{Evolution of the Hubble parameter $H(z)$ for flat ($\kappa=0$), closed ($\kappa=1$), and open ($\kappa=-1$) POQCO geometries compared against the Cosmic Chronometers dataset and the $\Lambda$CDM baseline.}
    \label{fig:cc_contours2}
\end{figure}

As illustrated in Figure~\ref{fig:cc_contours2}, these two curved geometries do not provide a statistically viable fit to the Cosmic Chronometers dataset. However, they reveal notable phenomenological features. Most significantly, the closed universe ($\kappa=1$) possesses a tightly constrained evolutionary trajectory that closely mimics a standard matter-dominated background, where the scale factor scales as $a(t) \propto t^{2/3}$. This scaling behavior is directly inherited from the dominance of the $(1+z)^{3/2}$ scaling envelope in Eq.~\eqref{eq:H_curvature} as the internal curvature modifications pull the root distribution of the auxiliary trajectory towards its asymptotic bound.

\end{appendix}


\begin{thebibliography}{}

\bibitem{Riess98}A. G. Riess et al, {\em Observational Evidence from Supernovae for an Accelerating Universe and a Cosmological Constant}, Astron. J. 116, 1009 (1998).

\bibitem{Perl99} S. Perlmutter  et al, {\em Measurements of $\Omega$ and $\Lambda$ from 42 High-Redshift Supernovae},   	Astrophys. J. 517, 565 (1999).




    

\bibitem{ABL} Y.  Aharonov,  P.  Bergmann,  and  J.  Lebowitz, {\em Time  Symmetry  in  the  Quantum  Process  of Measurement}, Phys. Rev. 134, B1410 (1964).

\bibitem{Ana25}C. Anastopoulos, {\em Final States in Quantum Cosmology: Cosmic Acceleration as a Quantum Post-Selection Effect},  Phys. Rev. D104, 064045 (2025).

\bibitem{twostate} Y. Aharonov and L. Vaidman, {\em The Two-State Vector Formalism of Quantum Mechanics: an Updated Review}, in `Time in Quantum Mechanics'', edited by J. G. Muga, R. Sala Mayato and I. L. Egusquiza (Springer, 2008).

 \bibitem{SGV} R. Silva, Y. Guryanova, N. Brunner, N. Linden, A. J. Short, and S. Popescu, {\em Pre- and Postselected Quantum States: Density Matrices, Tomography, and Kraus Operators}, 
Phys. Rev. A89, 012121 (2014).
 
\bibitem{Omn89}R. Omn\'es, {\em Logical Reformulation of Quantum Mechanics. IV. Projectors in
Semiclassical Physics}, J. Stat. Phys. 59, 223 (1989).

\bibitem{GeHa2} M. Gell-Mann and J.  B.  Hartle, {\em Classical Equations
for Quantum Systems},  Phys. Rev.   D47, 3345 (1993).


\bibitem{Ana23} C. Anastopoulos,{\em Quantum Theory: A Foundational Approach} (Cambridge: Cambridge University Press 2023).

 \bibitem{IsKa} C. J. Isham and A. C. Kakas, {\em A Group Theoretical Approach to the Canonical
Quantisation of Gravity. I. Construction of the
Canonical Group}, Class. Quantum Grav. 1 621 (1984).

\bibitem{Isham83} C. J. Isham, {\em Topological and global aspects of quantum theory}, in:  DeWitt, B. S., and Stora, R. (eds), ``Les Houches, Session XL: Relativity,
Groups and Topology II" (Amsterdam: North Holland, 1983).

\bibitem{EMCEE} 
D. Foreman-Mackey, D. W. Hogg, D. Lang, and J. Goodman, 
{\em emcee: The MCMC Hammer}, 
Publ. Astron. Soc. Pac. {\bf 125}, 306 (2013).

 

\bibitem{Scolnic22}
D. Scolnic, D. Brout, A. Carr, A. G. Riess, T. M. Davis, et al., 
{\em The Pantheon+ Analysis: The Full Data Set and Light-curve Release}, 
Astrophys. J. {\bf 938}, 113 (2022).

\bibitem{Akaike74}
H. Akaike, {\em A New Look at the Statistical Model Identification}, 
IEEE Trans. Autom. Control {\bf 19}, 716 (1974).

\bibitem{Schwarz78}
G. Schwarz, {\em Estimating the Dimension of a Model}, 
Ann. Statist. {\bf 6}, 461 (1978).

\bibitem{Liddle07}
A. R. Liddle, {\em Information Criteria for Astrophysical Model Selection}, 
Mon. Not. R. Astron. Soc. Lett.  377, L74 (2007).

 \bibitem{Yu18}
H. Yu, B. Ratra, and F.-Y. Wang, {\em Hubble Parameter and Baryon Acoustic Oscillation Measurement Constraints on the Hubble Constant, the Deviation from the Spatially Flat $\Lambda$CDM Model, the Deceleration–Acceleration Transition Redshift, and Spatial Curvature}, 
Astrophys. J. {\bf 856}, 3 (2018).

\bibitem{DaMa26} P. C. W. Davies and J. Magueijo, {\em Teleocosmology and Quantum Post-selection},  	arXiv:2606.02514.



\bibitem{ZWS} I. Zlatev,  L.-M, Wang, and P. J. Steinhardt,  {\em Quintessence, Cosmic Coincidence, and the Cosmological Constant},  Phys. Rev. Lett. 82, 896  (1999).

\bibitem{Visser}M. Visser, {\em Jerk, Snap and the Cosmological Equation of State},
Class. Quantum Grav. 21, 2603 (2004).


\bibitem{Omn1} R. Omn\'es, {\em The Interpretation of Quantum Mechanics} (Princeton: Princeton University Press 1994).



\bibitem{Kass95}
R. E. Kass and A. E. Raftery, {\em Bayes Factors}, 
J. Amer. Statist. Assoc.   90, 773 (1995).

\end{thebibliography}
\end{document}